\documentclass[prl, twocolumn]{revtex4-1}  
\usepackage{graphicx,graphics,color,epsfig} 
\usepackage{amsmath, amssymb,amsfonts} 
\usepackage{bm,times,xspace,mhchem} 
\usepackage[colorlinks,citecolor=blue,linkcolor=red]{hyperref} 

\begin{document}

\title{Quantum Spin of Elastic wave}

\author{Yang Long}
\author{Jie Ren}
 \email{Xonics@tongji.edu.cn}
 \author{Hong Chen}
\affiliation{%
Center for Phononics and Thermal Energy Science, China-EU Joint Center for Nanophononics, Shanghai Key Laboratory of Special Artificial Microstructure Materials and Technology,
School of Physics Sciences and Engineering, Tongji University, Shanghai 200092, China
}%

\date{\today}

\maketitle
{\bf Unveiling intrinsic spins of propagating waves usually offers people a fundamental understanding of the geometrical and topological properties of waves from classical to quantum aspects~\cite{lodahl2017chiral, Bliokh2015Transverse}. A great variety of research has shown that transverse waves can possess non-trivial quantum spins and topology without help of strong wave-matter interaction \cite{bliokh2015spin, aiello2015transverse, cardano2015spin}. However, until now we still lack essential physical insights about the spin and topological nature of longitudinal waves. Here, demonstrated by elastic waves we uncover unique quantum spins for longitudinal waves and the mixed longitudinal-transverse waves that play essential roles in topological spin-momentum locking. Based on this quantum spin perspective, several abnormal phenomena beyond pure transverse waves are attributed to the hybrid spin induced by mixed longitudinal-transverse waves. The intrinsic hybrid spin reveals the complex spin essence in elastic waves and advances our understanding about their fundamental topological properties. We also show these spin-dependent phenomena can be exploited to control the wave propagation, such as non-symmetric elastic wave excitation by spin pairs, uni-directional Rayleigh wave and spin-selected elastic wave routing. These findings are generally applicable for arbitrary waves with longitudinal and transverse components. }

Almost 150 years ago, Helmholtz's theorem has unveiled the fact of fundamental geometry 
that any vector field can be decomposed into two parts $\bm{u} = \bm{u}_{L} + \bm{u}_{T}$~\cite{Helmholtz1958}, a curl-free component 
and a divergence-free component
, namely, longitudinal wave and transverse wave, respectively. Since then tons of studies have shown that the transverse propagating waves, i.e. optical waves, carry their characteristic spin determined by the wave vector and polarization profile~\cite{Bliokh2015Transverse,lodahl2017chiral}. Due to the essential geometrical and topological properties, non-trivial Berry phase and quantum spin Hall effect (QSHE) can be induced and possessed for transverse waves even in accutum~\cite{Bliokh2015OPTICS,Stone2015Topology,VanMechelen16, Bliokh2014Extraordinary,Bekshaev2015Transverse,PhysRevB.74.174302,Bliokh2008Geometrodynamics}. Experiments also have shown the spin of transverse waves can supply a robust and powerful approach to control the wave flow~\cite{Yin2013Photonic, Lin2013Polarization, Lin2013Polarization,Shitrit2013Spin,sala2015spin}. Nevertheless, despite of distinct geometrical properties, no studies have yet been made to discuss the corresponding physical characteristics of longitudinal waves, especially their quantum spin and topological nature.

\begin{figure}[!tp]
\centering
\includegraphics[width=\linewidth]{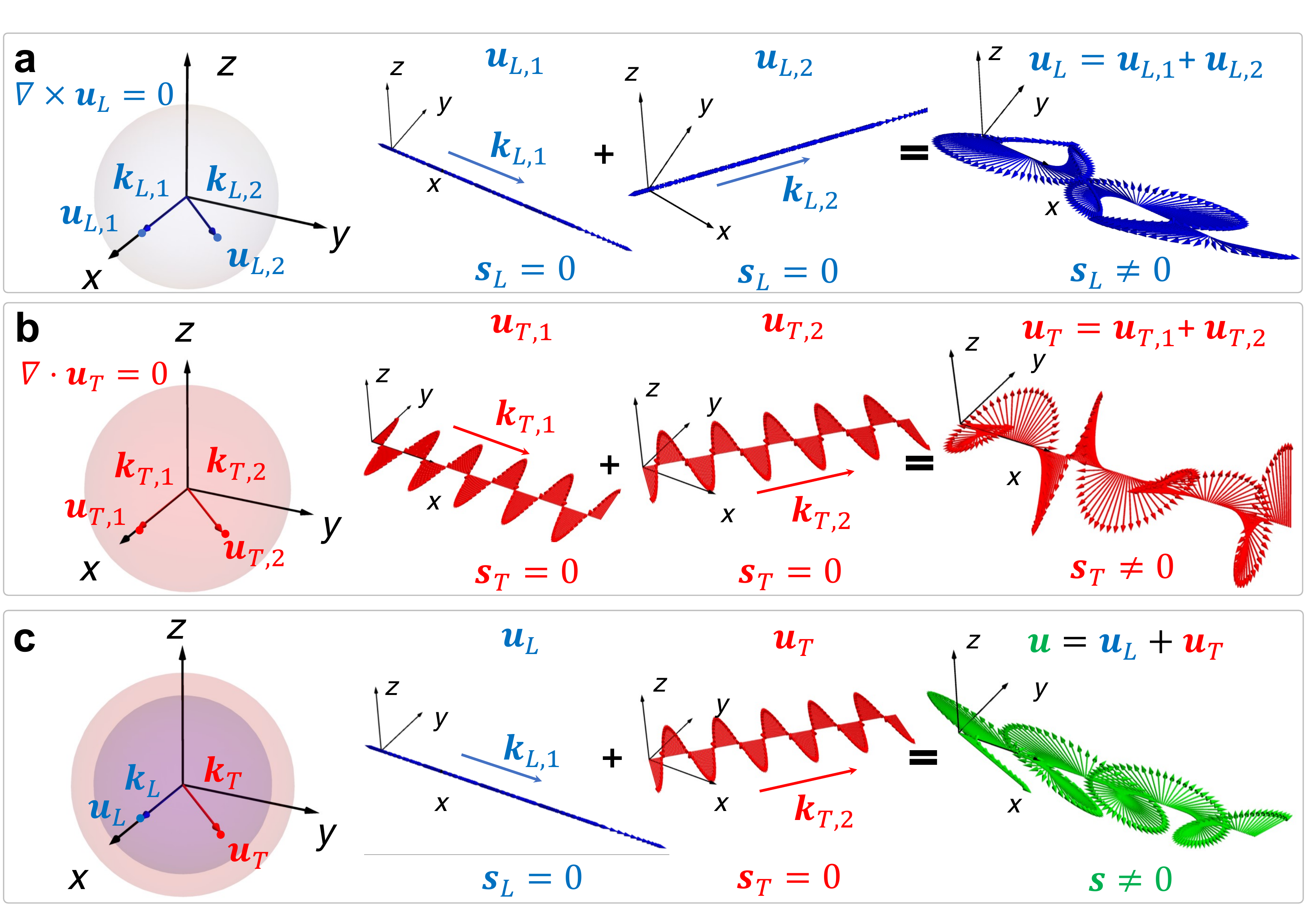}
\vspace{-8mm}
\caption{{\bf Geometry, topology and spin of elastic wave}. The strong relations between polarization profile and wave vector reflect the intrinsic spin orbital couplings: $\bm{k}_L\times\bm{u}_L=0$ and $\bm{k}_T\cdot\bm{u}_T=0$, {\it i.e.}, the spin-momentum locking. {\bf a}, For the combinations of two longitudinal waves with different wave vector $\bm{k}_{L,1} \neq \bm{k}_{L,2}$, $\bm{u}_L = \bm{u}_{L,1} + \bm{u}_{L,2}$, the total elastic field carries non-trivial spin angular momentum density $\bm{s}_L\neq 0$ due to the wave interference. {\bf b}, The transverse waves in same settings, $\bm{u}_T = \bm{u}_{T,1} + \bm{u}_{T,2}$,  also induce non-trivial spin density $\bm{s}_T \neq 0$. {\bf c}, For the total elastic waves that contain longitudinal and transverse plane waves simultaneously, $\bm{u} = \bm{u}_{L} + \bm{u}_{T}$, the total elastic spin is attributed to the hybrid spin $\bm{s}=\bm{s}_h$, which reflects the major geometrical difference between longitudinal and transverse waves. The spheres in figure are $\bm{k}$-spheres and the real parts of displacement fields are plotted.}
\label{fig:geometryandtopology}
\end{figure}

The elastic wave describes the basic dynamic principle of how solid objects deform and become internally stressed in a periodic form~\cite{Auld1973Acoustic}, which can reflect the properties from classical solid motion to lattice oscillation in quantum field. Distinct from the transverse optical wave and longitudinal acoustic wave, the elastic wave can support both the longitudinal and transverse components simultaneously~\cite{Auld1973Acoustic}. Thus elastic wave serves as the ideal platform for exploring quantum topological properties of longitudinal waves, as well as the intrinsic interaction between longitudinal and transverse ones. In this Letter, we find that the longitudinal component possesses its own unique spin density that plays an important role in geometrical and topological properties of elastic wave. The mixed longitudinal-transverse wave will carry an extra spin due to their intrinsic hybridization, which will be responsible for abnormal phenomena beyond pure transverse waves. By analyzing the quantum spin of elastic wave (general for other vector field) in detail, we identify that the total spin angular momentum density should be dissected into three contributions, as (see Supplementary):
\begin{equation}
\bm{s} = \bm{s}_L + \bm{s}_T + \bm{s}_{h}
\label{eq:totalspin}
\end{equation}
where $\bm{s}_L = \langle\bm{u}_L|\hat{\bm{S}}|\bm{u}_L\rangle \propto {\rm Im}[\bm{u}^*_L\times\bm{u}_L]$ is the spin contribution from longitudinal wave $\bm{u}_L$, $\bm{s}_T = \langle\bm{u}_T|\hat{\bm{S}}|\bm{u}_T\rangle \propto {\rm Im}[\bm{u}^*_T\times\bm{u}_T]$ is from transverse wave $\bm{u}_T$, and specially $\bm{s}_h = \langle\bm{u}_L|\hat{\bm{S}}|\bm{u}_T\rangle + \langle\bm{u}_T|\hat{\bm{S}}|\bm{u}_L\rangle \propto {\rm Im}[\bm{u}^*_T\times\bm{u}_L]$ is the hybrid spin due to non-trivial spin projections among these two states, where $\hat{\bm{S}}$ is the quantum spin operator for elastic wave. 
Spin is one of the most important physical properties in quantum mechanics and corner stone for topological states, which can reveal the complex intrinsic interactions among multiple physical mechanisms.
Except the well known spin $\bm{s}_T$~\cite{Berry2009Optical, Bliokh2015Transverse, lodahl2017chiral}, $\bm{s}_L$ and $\bm{s}_h$ are hidden and unobservable in pure transverse wave. We will fill this gap and advance the understanding about the geometrical and topological nature of elastic wave, as well as arbitrary vector field.

\begin{figure}[!tp]
\centering
\includegraphics[width=\linewidth]{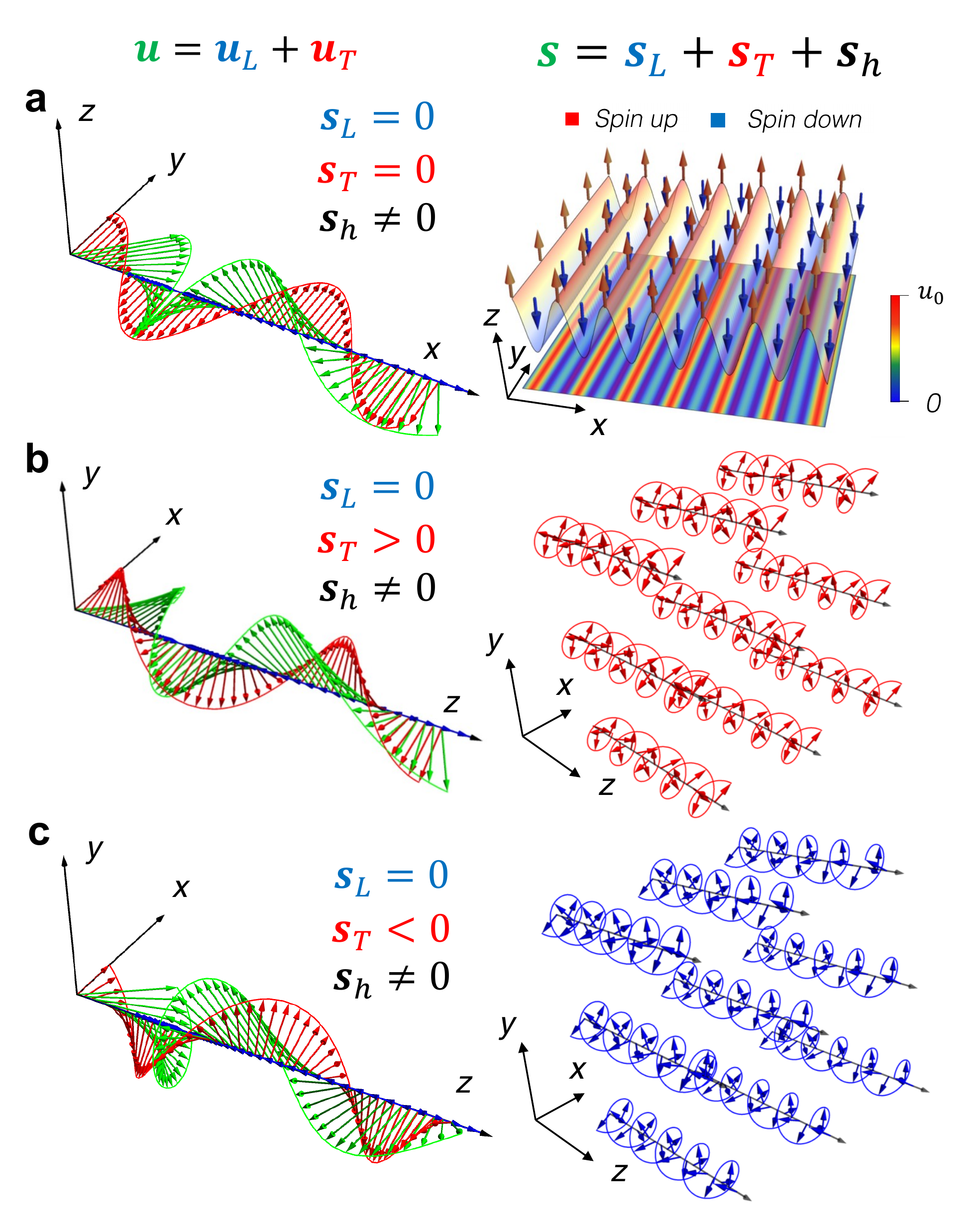}
\vspace{-8mm}
\caption{{\bf Quantum spin hybridization of elastic wave}. The left figures are plots about real parts of displacement fields and the right figures are corresponding spin textures. {\bf a}, When two kinds of elastic waves propagate along same direction, longitudinal wave $\bm{u}_L \propto \exp(ik_L x) \bm{e}_x$ and transverse wave $\bm{u}_T \propto \exp(ik_T x)\bm{e}_y$ in plane wave form, the total elastic wave $\bm{u}=\bm{u}_L + \bm{u}_T$ reveal the intrinsic spin hybridization raised from the geometrical and topological differences. The total spin varies periodically along the propagation direction $x$, $\bm{s} = \bm{s}_{h} \propto \sin((k_L-k_T)x) \bm{e}_z$. The $z$ component of $\bm{s}$ is plotted beyond the total displacement field. Another intriguing case about elastic spin is rotating spin texture. {\bf b}, For the combination of longitudinal wave and circularly-polarized transverse wave with non-trivial spin $\bm{s}_T >0$, the total elastic spin becomes clockwise rotating along the propagation direction. {\bf c}, When $\bm{s}_T <0$, the total elastic spin becomes anti-clockwise rotating along the propagation direction. $\bm{s} \propto \mp\cos((k_L-k_T)x) \bm{e}_x - \sin((k_L-k_T)x)\bm{e}_y \pm \bm{e}_z$, + for $\bm{s}_T > 0$ and - for $\bm{s}_T < 0$, respectively.}
\label{fig:hybridspin}
\end{figure} 

Considering the elastic waves in homogeneous isotropic solid, the linear elastic equation can be decoupled into two independent forms: longitudinal wave $\bm{u}_L$ and transverse wave $\bm{u}_T$~\cite{Auld1973Acoustic}. The propagating elastic waves in solid are polarized plane waves. With the basic geometrical conditions of longitudinal wave $\nabla\times\bm{u}_L=0$ and transverse wave $\nabla\cdot\bm{u}_T=0$, their polarization vectors will become momentum dependent in plane wave form. The transverse wave can hold circularly polarized propagating form and carry transverse wave spin naturally $\bm{s}_T \neq 0$ ~\cite{Bliokh2015Transverse, Berry2009Optical}. For longitudinal wave, there is no extra freedom to constitute such non-trivial chirality $\bm{s}_L = 0$. But geometrically, we can use two wave interferences~\cite{Bekshaev2015Transverse} to induce effective ``circularly polarized'' profile for total wave field in Fig.~\ref{fig:geometryandtopology}. For the combination of two waves with different wave vectors, the total elastic wave will possess the similar property with circularly polarized wave profile. Specially, the total elastic wave will keep its original geometrical characteristic conditions, namely, $\nabla\times\bm{u}_L = 0$ in Fig.~\ref{fig:geometryandtopology}a, $\nabla\cdot\bm{u}_T = 0$ in Fig.~\ref{fig:geometryandtopology}b, but with nonzero spin $\bm{s}_L\neq0$ and $\bm{s}_T\neq0$, respectively. Besides these conventional cases, there exists one special spin form $\bm{s}_h$ in the mixed longitudinal-transverse wave in Fig.~\ref{fig:geometryandtopology}c. Although wave interference only occurs in waves of same type conventionally, the longitudinal-transverse wave mixing can be considered as generalized effective wave interference to produce the new non-trivial hybrid spin: $\bm{s}_h$.

\begin{figure}[!tp]
\centering
\includegraphics[width=0.8\linewidth]{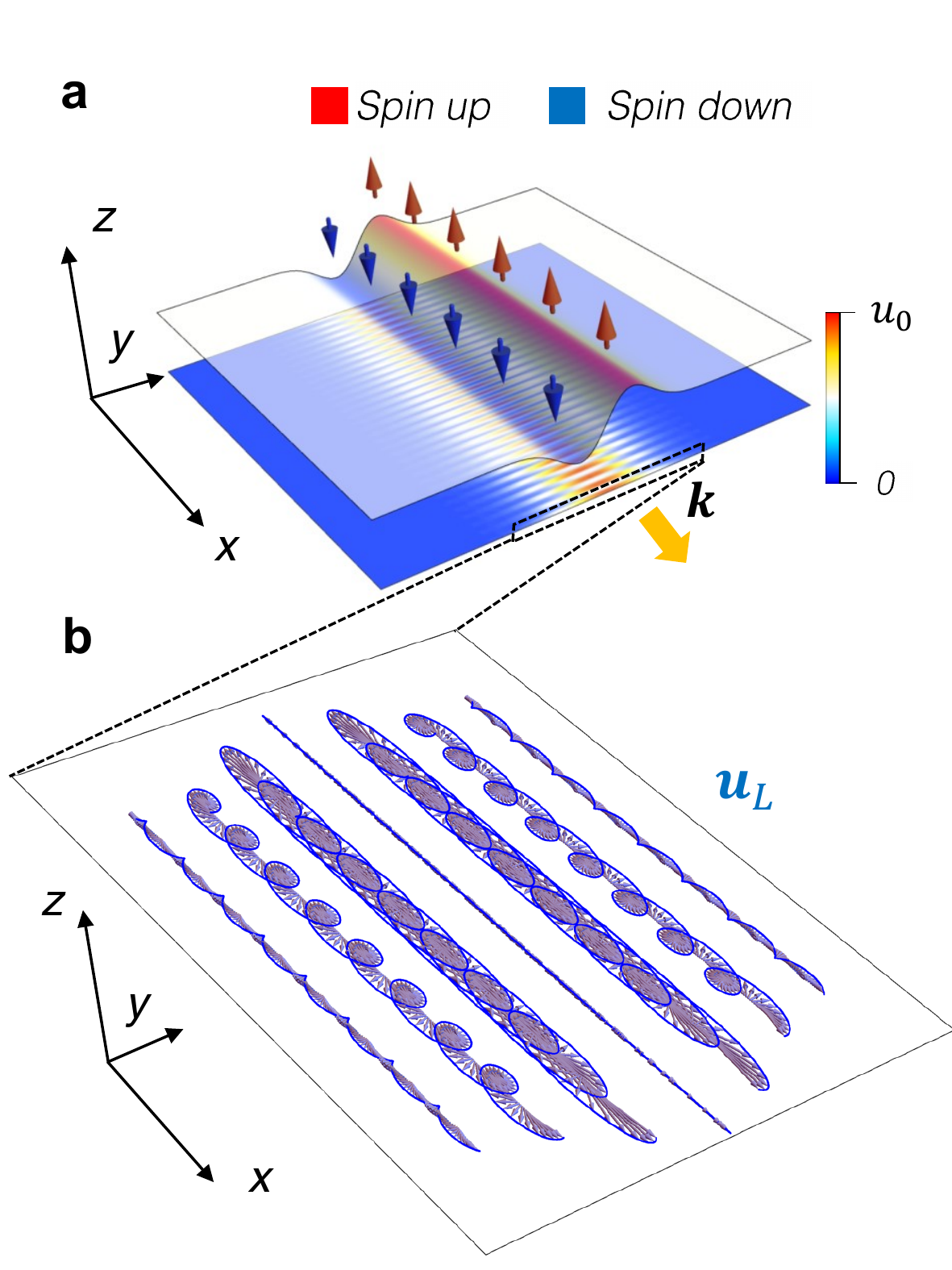}
\vspace{-5mm}
\caption{{\bf Non-trivial spin texture in elastic longitudinal wave beam}. Besides the wave interference approaches, the elastic Gaussian wave beam can also induce non-trivial spin density distribution. {\bf a}, For the longitudinal wave Gaussian beam in $xOy$ plane, $|\bm{u}_L| \propto \exp(-y^2/\delta^2)$, it will carry the anti-symmetric spin density around the center of beam, $\bm{s} \propto \frac{4y}{k_L \delta^2}\exp(-2y^2/\delta^2)\bm{e}_z$. The $z$ component of $\bm{s}$ is plotted above the displacement field of longitudinal wave Gaussian beam. {\bf b}, The real part of displacement field behaves effectively as circularly polarized wave profile, whose rotation reflects the non-trivial transverse spin properties along both sides of Gaussian beam.}
\label{fig:gaussianbeam}
\end{figure}

Topological properties are associated with the momentum dependent polarization profile, especially spin states~\cite{Kane2005Z2,Moore2010The}. Considering the basic momentum dependent geometrical conditions of elastic wave, we can see that the polarization profile of $\bm{u}_L$ ($\bm{u}_T$) is normal (tangent) to the $\bm{k}$-sphere. Actually, these essential geometrical relations underlie the strong spin-momentum locking of elastic waves in isotropic homogeneous solid, which is reminiscent of the spin of light in vacuum without strong matter-wave interactions~\cite{Bliokh2015OPTICS,bliokh2015spin}. The non-trivial Berry curvature and non-zero spin Chern number have been proposed theoretically~\cite{PhysRevB.74.174302,Bliokh2015OPTICS} in the circularly polarized transverse wave  and verified experimentally~\cite{Neugebauer2015,guo2017photonic}. Similarly, here the spin-momentum locking of longitudinal wave also implies its unique topological nature, but with trivial Berry curvature and spin Chern number. These topological invariant difference defined by distinct geometrical relations will become the main reason behind intrinsic spin hybridization in mixed longitudinal-transverse wave. 

\begin{figure}[!tp]
\centering
\includegraphics[width=0.8\linewidth]{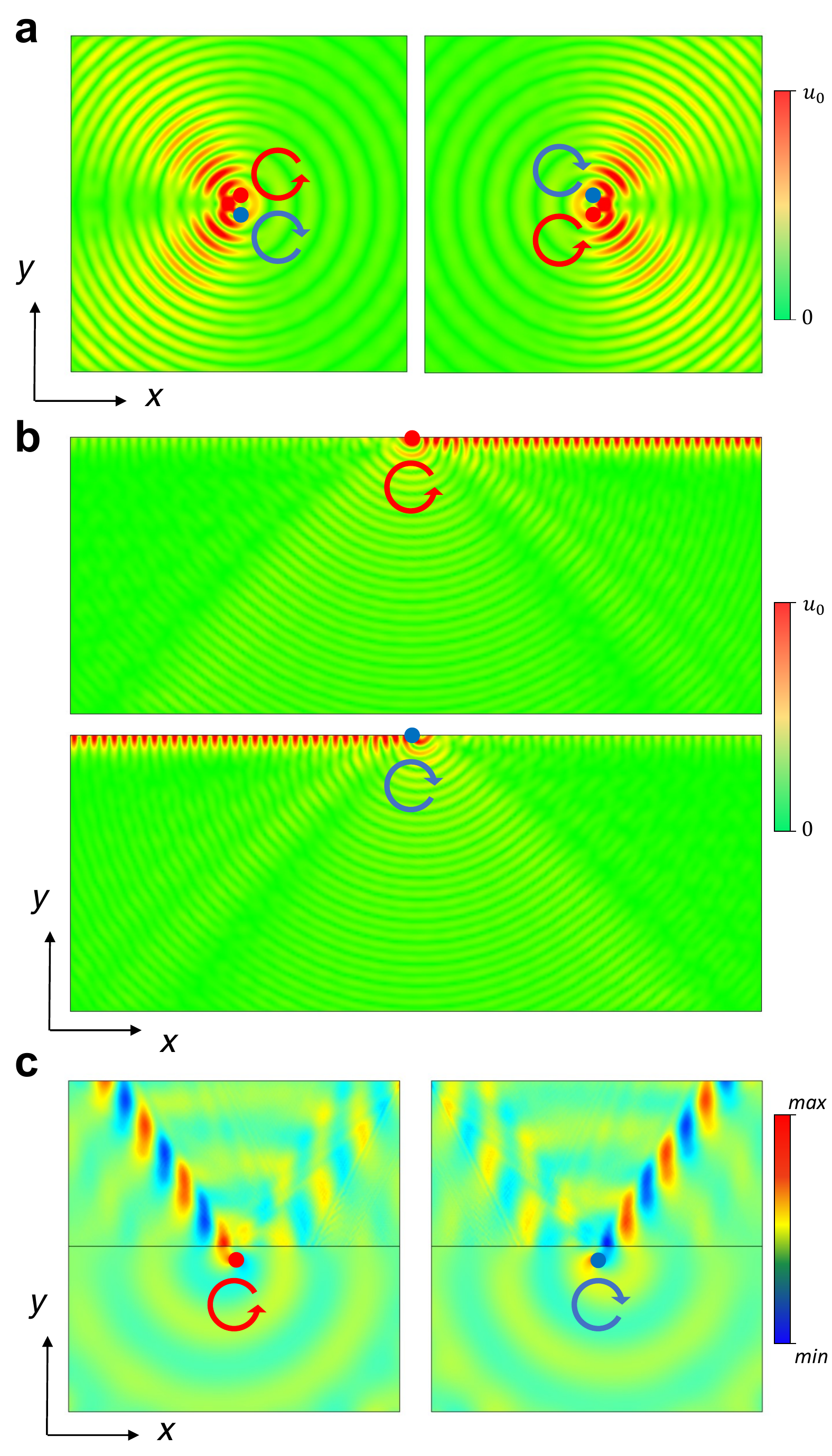}
\vspace{-5mm}
\caption{{\bf Spin momentum locking in elastic wave}. The strong anisotropic excitation of spin-selected elastic wave can be realized based on the spin-momentum locking. {\bf a}, Non-symmetric elastic wave excitation in bulk can be achieved by exploiting two circularly-polarized elastic loads to induce effective opposite spin pair. The total displacement field is plotted. {\bf b}, Unidirectional elastic surface mode, Rayleigh wave, can be excited selectively by setting different circularly polarized elastic loads. The total displacement field is plotted. {\bf c}, The spin controlled elastic wave routing in bulk can be realized in anisotropic materials. The area above black line is anisotropic material and the other area below is isotropic material. The divergence of total displacement field that reflects the longitudinal component is plotted. The spin-momentum locking analysis and calculation details can be found in Supplementary.}
\label{fig:spinmomentumlocking}
\end{figure}

To exemplify the intrinsic hybrid spin $\bm{s}_h$, we consider the simplest plane wave form  in Fig.~\ref{fig:hybridspin}a. A longitudinal elastic wave and the other linear polarized transverse wave propagate in parallel. They both individually carry zero spin density but the complex geometrical form of their total elastic field reflects the intrinsic hybridization between them. This intrinsic hybridization induces non-zero hybrid spin $\bm{s}_h \neq 0$, which contributes to the total spin $\bm{s}=\bm{s}_h$. Importantly, this hybrid spin varies periodically along propagating direction due to spin-momentum locking: $\bm{s}_h \propto \sin((k_L-k_T)x)$. This fluctuation induces non-zero spin momentum density $\bm{p}^s\equiv\frac{1}{2}\nabla\times \bm{s} \neq 0$, which is observable and detected in experiments. The hybrid spin will introduce some interesting spin phenomena, such as wave spin lattice (see Supplementary) and rotating spin texture. Considering the combination of longitudinal wave and circularly polarized transverse wave with non-trivial spin $\bm{s}_T \neq 0$ in Fig.~\ref{fig:hybridspin}b and c, one can find the total spin direction will rotate along the propagating direction. The rotating direction is associated with the spin profile of transverse wave: clockwise for $\bm{s}_T >0$ and anti-clockwise for $\bm{s}_T<0$. Conventionally, $\bm{p}^s$ for the circularly polarized transverse wave vanishes due to the cancellation of neighbour spin currents~\cite{Bliokh2014Extraordinary}. But mixed with longitudinal wave, the total elastic wave will carry detectable non-zero $\bm{p}^s \neq 0$ with help of nonzero hybrid spin $\bm{s}_h$.

Besides two wave interference, the pure longitudinal wave can also carry non-trivial spin density in spatial confined Gaussian wave form. We calculate the spin density of elastic longitudinal wave in Fig.~\ref{fig:gaussianbeam}a~\cite{Neugebauer2015}. One can see that the two sides of longitudinal wave beam have opposite spin distribution, which can be described by the non-zero spin density of spatial Gaussian decay field $\bm{s} \propto \frac{4y}{k_L \delta^2}\exp(-2y^2/\delta^2)\bm{e}_z$, where $\delta$ is the width of Gaussian decay. The non-trivial spin originates from the effective imaginary vector field along Gaussian decay direction that induces the circularly polarized wave profile, with the geometrical constraint of longitudinal wave $\nabla\times\bm{u}_L = 0$. In Fig.~\ref{fig:gaussianbeam}b, we indeed see that the real part of displacement fields form opposite circularly polarized wave profile on both sides of the Gaussian beam, which can carry non-trivial  spin along $z$ similar to the case of transverse waves~\cite{Neugebauer2015,Bliokh2015Transverse}. This anti-symmetric spin texture underlies the strong spin-momentum locking in Gaussian beam. Introducing the generalized longitudinal wave Gaussian beam solution $\bm{u}_{L,G}(\bm{k})$, one can calculate its Berry connection $\mathcal{A} = -i \int_{Full} \bm{u}_{L,G}^*(\bm{k})\cdot(\nabla_{\bm{k}})\bm{u}_{L,G}(\bm{k})d\bm{s}$, Berry curvature $\mathcal{F} = \nabla_{\bm{k}}\times \mathcal{A}$, and Chern number $C = \frac{1}{2\pi} \oint \mathcal{F} d^2\bm{k}$ for the full space. In this case, it will result trivial Berry curvature $\mathcal{F} = 0$ and zero Chern number $C=0$. Considering that the full space integral of spin density will vanish due to anti-symmetric texture, one can adjust the integral only for half space, which can induce non-trivial Berry curvature $\mathcal{F}_\pm \neq 0$ and non-zero topological invariants $C_\pm = \pm 1$, with $\pm$ meaning the integral in opposite half spaces. The difference of them, $\Delta C = C_{+} - C_{-} = 2$ will be a well defined topological number~\cite{Delplace1075} for longitudinal wave Gaussian beam to describe its spin-momentum locking relation.

Strong spin texture momentum locking in elastic wave inspires us a new scheme to excite spatial non-symmetric elastic wave. We exploit two different circularly polarized elastic loads to serve as the pair of spin sources to induce corresponding spin texture in Fig.~\ref{fig:spinmomentumlocking}a. We can see that the propagating direction of excited elastic waves performs strong dependence on the spin pair profile. Besides this scheme, we can excite the unidirectional Rayleigh wave and specially associate its propagation direction with spin profile, by using circularly elastic loads on boundary of solid in Fig.~\ref{fig:spinmomentumlocking}b, reminiscent of the phononic QSHE~\cite{PhysRevLett.115.104302}. Meanwhile, in anisotropic material, we can also realize spin-selected elastic routing in Fig.~\ref{fig:spinmomentumlocking}c, which can be regarded as QSHE in bulk~\cite{Kapitanova2014Photonic}. The quantum spin of elastic wave is responsible for all of these phenomena. 



Our work unveils the physical essence and topological nature of the intrinsic quantum spin in elastic wave. Moreover, it gives a perspective on how to fill the gap in the topological quantum theory about arbitrary vector fields. Potentially, our results could be extended to connect other ``spin''-related systems from conventional elastic wave devices~\cite{PhysRevLett.114.114301} to nano-scale phononic materials~\cite{maldovan2013sound}, e.g., the high efficient spin-selected wave emitter,  spin-sensitive detector, or high-speed multi-channel information transfer. Our results could advance the fundamental understanding about spin and topological properties for general waves and serve as a powerful routine for the realization of quantum wave-spin computation network in the future.

{\bf Acknowledgements} This work is supported by the National Natural Science Foundation of China (Grant Nos. 11775159 and 11234010), the National Key Research Program of China (No. 2016YFA0301101), and the National Youth 1000 Talents Program in China.

{\bf Author Contributions} JR conceived the project. YL carried out all the numerical calculations and figure plots. YL and JR derived the theory and developed the physical analysis. YL wrote the Supplementary Information. YL, JR, and HC all contributed to discussion, interpreting the data and the writing.

{\bf Competing Interest} The authors declare that they have no competing financial interests.

{\bf Correspondence} Correspondence and requests for materials should be addressed to Jie Ren (email: xonics@tongji.edu.cn).


\end{document}